\documentclass{elsarticle}

\makeatletter
\def\ps@pprintTitle{%
 \let\@oddhead\@empty
 \let\@evenhead\@empty
 \def\@oddfoot{\centerline{\thepage}}%
 \let\@evenfoot\@oddfoot}
\makeatother
\usepackage{graphicx}
\usepackage{caption}
\usepackage{subcaption}
\usepackage[misc,geometry]{ifsym}
\graphicspath{{./fig/}}
\usepackage{amssymb}

\begin{document}

\begin{frontmatter}

\title{Fog enabled distributed training architecture for federated learning}

% \address{Radarweg 29, Amsterdam}
% \fntext[myfootnote]{Since 1880.}

% \author[mymainaddress]{Aditya Kumar}
\author{Aditya Kumar}

\author{Satish Narayana Srirama\corref{mycorrespondingauthor}}
\cortext[mycorrespondingauthor]{Corresponding author}
\ead{satish.srirama@uohyd.ac.in}

\address{Cloud \& Smart Lab,  SCIS\\ University of Hyderabad\\ Telangana, India}

\begin{abstract}
The amount of data being produced at every epoch of second is increasing every moment. Various sensors, cameras and smart gadgets produce continuous data throughout its installation. Processing and analyzing raw data at a cloud server faces several challenges such as bandwidth, congestion, latency, privacy and security. Fog computing brings computational resources closer to IoT that addresses some of these issues. These IoT devices have low computational capability, which is insufficient to train machine learning. Mining hidden patterns and inferential rules from continuously growing data is crucial for various applications. Due to growing privacy concerns, privacy preserving machine learning is another aspect that needs to be inculcated. In this paper, we have proposed a fog enabled distributed training architecture for machine learning tasks using resources constrained devices. The proposed architecture trains machine learning model on rapidly changing data using online learning. The network is inlined with privacy preserving federated learning training. Further, the learning capability of architecture is tested on a real world IIoT use case. We trained a neural network model for human position detection in IIoT setup on rapidly changing data.
\end{abstract}


\end{frontmatter}

%\linenumbers

\section{Introduction}

With advances in digital technology, Internet of Things (IoT)~\cite{GUBBI20131645} devices are prevailing everywhere. Multiple sensors, cameras, mobiles, and smart gadgets are installed to provide support in decision making. As technology progresses, the reliance on such devices is increasing day by day. Deployment of various IoT devices has increased exponentially nowadays.  The devices include simple sensors to very sophisticated industrial tools that exchange data/information through the internet. In the past few years, the number of IoT devices has increased rapidly. Currently, there are more than 10 billion IoT devices available worldwide, which is expected to be around 17 billion in 2025 and 26 billion by 2030~\cite{holst_2021}. Every standalone device produces data which is shared with other devices for further processing. The IoT devices placed at the edge layer are generally resource constrained. However, devices such as cameras or sensors generate continuous data by sensing the environment. These devices have very crucial information to mine that can be used to achieve a business goal. As the number of devices increases, the velocity and volume of data increases significantly over time.  Processing and analysis of continuously generated data from resource constrained remote devices is a challenging task. Training a machine learning(ML)model on such large distributed data can be used to solve real world computational problems.

In the conventional machine learning paradigm, the training is done at the central server/cloud. The data is generated at the edge device, which is sent to the cloud. The cloud stores all data and performs training. The cloud-IoT architecture consumes large bandwidth while transferring raw data to the cloud that also creates network congestion~\cite{chang2018internet}. The high latency is an issue in the cloud-IoT model that limits it for continuous learning. Additionally, data privacy concern is another major challenge in the data collection at the server. Fog computing~\cite{Bonomi10.1145,buyya2019fog} brings computational resources closer to the edge nodes that can efficiently process the raw data. Various data generating devices closely communicate with the nearest fog node for local computation. A fog node has enough computational capacity to process periodically collected data. The fog node can manoeuvre the associated IoT devices and directly communicates to the central server for further knowledge discovery. The distributed machine learning can be used to learn the hidden patterns from raw data efficiently. Federated learning trains a machine learning model from distributed data without sharing raw data to the server.   

Distributed machine learning is a  multi-node training paradigm where a participating node trains its model and collaborates with each other or the server for optimization. Federated learning~\cite{pmlr-v54-mcmahan17a} training proposed by McMahan et al. is a decentralized machine learning technique that can train an artificial neural network (ANN) without sending and storing raw data at the server. The algorithm trains a global model in collaboration with various devices without sharing data. Every participating device trains a model locally on their local data. The central curator coordinates with all the devices to create a global model. The locally trained model parameters such as weights and biases of ANN are shared with the central server rather than raw data. The server further aggregates model parameters from participating devices and creates a global model. This iterative process continues till the convergence of the model. In the entire training, the raw data is never shared with anyone that makes the system overall privacy preserving. Federated learning is applied to various tasks such as smart city, autonomous driving cars, industrial automation, etc. The data generating IoT devices are resources constrained that cannot train a machine learning model on the edge. Whereas, a fog computing paradigm brings computational efficiency near to IoT devices that can directly participate in federated learning. A fog enabled cloud-IoT model has the potential to quickly process continuous data. 

In this paper, we have proposed an IoT-fog-cloud architecture to train a neural network on continuously generated distributed data. The paper tries to combine fog computing and federated learning for model creation. This addresses online training of continuously generating data using resource constrained devices. The proposed architecture is shown in Fig~\ref{fig:architecture}. The IoT devices at the edge layer generate continuous data and share it with the fog node. The fog node is capable of online training that trains models in collaboration with the central cloud. The central server applies federated learning with the fog layer. The fog nodes capture periodic data from their associated IoT devices. It performs local model training and communicates with the cloud for global model optimization. The raw data generated at IoT devices are only shared with the local/nearest fog node. The fog layer is equipped with finite computational and storage resources that can process the raw data. The continuous data gets accumulated at fog layer/backup storage, whereas the training is done on a periodic/recent dataset only. Once the data is used for single shot training, it is not used for further training. The data does not leave the premises of fog architecture that makes the system more privacy sensitive. However, the stored data can be used for future references or it can be sent to the server if required by the application in performing long-term big data analytics. We have simulated the proposed architecture to train a neural network for safe position detection in real world Industrial Internet of Things (IIoT) setup using docker. The contributions of the paper are summarized in the following:

\begin{itemize}
    \item Proposed a fog enabled distributed training architecture for machine learning tasks. A hybrid of Fog computing and Federated learning paradigm is used for the model training.
    \item Online continuous training is done with rapidly changing/growing datasets.  The training includes only recent periodic data for modelling. The system assures privacy by restricting the raw data to the fog level. In addition, by not sharing raw data directly to the server, the system optimizes network bandwidth and congestion. 
    \item We simulated the proposed architecture with Docker container. To test the learning capability of the model, we used radar data to train safe position classification in Human Robot (HR) workspace. 
\end{itemize}

Rest of the paper is organized as follows. Section \ref{sec:relatedWork} discusses existing related articles. The proposed architecture and decentralized machine learning are discussed in \ref{sec:DFL}. Experimental setup and numerical results are shown in section \ref{sec:Evaluation}. Section \ref{sec:conclusion} concludes the paper with scope for future work.

\section{Related Work}
\label{sec:relatedWork}

Data processing and machine learning need huge computational and storage resources to execute a specific task. Multiple IoT devices have generated and continue to generate voluminous data. One of the efficient ways to achieve such a task is to rent a cloud computing facility. With virtually infinite resources, the cloud executes complex model training on big data. The cloud-IoT faces various challenges such as Bandwidth, Latency, Uninterrupted, Resources-constraint and security~\cite{chang2018internet}. The IoT devices are resource constraint which are connected through a wireless network to the cloud. These shortcomings obstruct smooth execution of the various tasks, specifically the real time processing. Fog computing proposes an alternative to the cloud that brings resources closer to IoT devices. It ensures low latency, network congestion, efficiency, agility and security~\cite{openfog}. This enabled efficient data processing at the fog layer and opens door to various applications such as smart cars, traffic control, smart buildings, real time security surveillance, smart grid, and many more. The fog layer has sufficiently enough resources to store and process raw data. This gives an advantage over a cloud to processing and decision making locally.

Bonomi et al.~\cite{Bonomi10.1145} have discussed about the role of fog computing in IoT and its applications. Fog computing provides localization that acted as a milestone in delay sensitive and real time applications. Data analytics on real time data have various applications based on context. Some of them, such as detection or controlling, need quick response typically in milliseconds or sub seconds, whereas other applications like report generation, global data mining tasks are long term tasks. Fog computing and cloud computing can performs interplay operations to achieve big data solutions. Fog responds to the real time processing task, which can be geographically distributed, and cloud computing proceed with big data analysis or knowledge consolidation. Due to proximity, fog computing is beneficial for delay sensitive applications, but it may lose importance when it gets congested. The number of jobs received at a particular fog node at a specific time may be higher, which cannot be processed quickly due to limited resources.  Al-khafajiy et al.~\cite{reading90118} have proposed a fog load balancing algorithm to request offloading that can potentially improve network efficiency and minimize latency of the services. The fog nodes communicate with each other to share their load optimally, which improves the quality of services of the network. Similarly, Srirama et al.~\cite{srirama2021akka}, have studied utilizing fog nodes efficiently with distributed execution frameworks such as Akka, based on Actor programming model. In this context, it is also worth mentioning that scheduling of applications/tasks in cloud and fog has been studied extensively in the last few years, which is summarized in the related work of Hazra et al.~\cite{hazra2020joint}.  

The Cloud-fog computing paradigm is more efficient, scalable and privacy preserving for machine learning model training. So edge centric computing framework is needed to solve various real time operations. Munusamy et al.~\cite{9512520} have designed a blockchain-enabled edge centric framework to analyze the real time data in Maritime transportation systems. The framework ensures security and privacy of the network and exhibits low latency and power consumption. Distributed machine learning training offers parallel data processing over the edge of the network. Kamath et al.~\cite{7842181} propose a decentralized stochastic gradient descent method to learn linear regression on the edge of the network. The work utilizes distributed environment to train regression model using SGD. The method process data at device level and avoids sending it to the cloud. Federated learning is another decentralized ML technique that trains models using very large set of low resourced participating devices~\cite{pmlr-v54-mcmahan17a,DBLP:journals/corr/KonecnyMRR16}. The proposed federated method collaborates with various participating devices to create a global ML model on decentralized data. Federated learning is done on low constraint devices that need efficient training strategies for uplink and down ink communication. Konečný et al.~\cite{1610.05492}, talks about efficient communication between cloud and devices. The authors have suggested sketched and structured updates for server communication that reduce the amount of data sent to the server.

The fog-cloud architecture is well suited for distributed machine learning training. Li et al.~\cite{9151163} have used cloud-fog architecture for secure and privacy preserving distributed deep learning training. The local training is given at the fog layer then it coordinates with the cloud server for aggregation. Additionally, it uses encrypted parameters and authentication of valid fog node to ensure legit updates. The central node works like a master node for information consolidation. It synchronizes the training from various devices. Due to stragglers or mobility of devices such as vehicles, drone the synchronous update creates difficulty in training.   Lu et al.~\cite{8843942} proposes asynchronous federated training for mobile edge computing.  The training is done similar to federated learning, but global model aggregation is done asynchronously. To ensure the privacy and security of the shared model, it adds adds noise to the parameters before sending it to the server. Luo et al.~\cite{9127160} have proposed a hierarchical federated edge learning framework to train low latency and energy-efficient federated learning. The framework introduces a middle layer that partially offloads cloud computational work. The proposed 3-layered framework aggregates model parameters at both fog layer and cloud layer while training is done at the remote device. Fog enabled federated learning can facilitate distributed learning for delay-sensitive applications. Saha et al.~\cite{9302575} proposed fog assisted federated learning architecture for delay-sensitive applications. The federated learning is done between edge and fog layer, then the central node heuristically steps in for global model aggregation. This training is done on geographically distributed network that optimizes communication latency by 92\% and energy consumption by 85\%. Most of the research assumes that data generating IoT devices contain enough computational resources to train ML models. Additionally, these IoT devices participate in distributed training with the complete dataset. Our proposed architecture is a three layered network for machine learning training. The edge layer generates raw data only, and the cloud layer consolidates the global model. Whereas, the fog layer participates in decentralized machine learning training with the central server. The federated training is done on continuously changing dataset generated by the edge layer.   

\section{Decentralized federated learning}
\label{sec:DFL}

Decentralized federated learning aggregates locally trained models on the central server. A global model is created by combining multiple independently trained models. The conventional machine learning approaches collect possible data $D$ to the server; then it learns a machine learning model $M$ using a sophisticated algorithm. In contrast, federated learning trains its model without collecting all possible data to the central server. It is a collaborative learning paradigm where $k$ number of participating devices trains a local model on their data $D_i$. Each participating device i contains their personal data $D_i$. The total data set is $D=\sum\limits_{i=1}^k D_i$. Here, $D_i$ is a collection of input data samples $(x_j, y_j)_{j=1}^n$ for supervised learning. Where $x_j \in \mathbb{R}^d$ is a $d$ dimensional input data and $y_j\in\mathbb{R}$ is the associated label for input $x_j$.  The device data $D_i$ are remotely generated or centrally distributed. 

\begin{figure}[h]
\centering
\includegraphics[width=\textwidth]{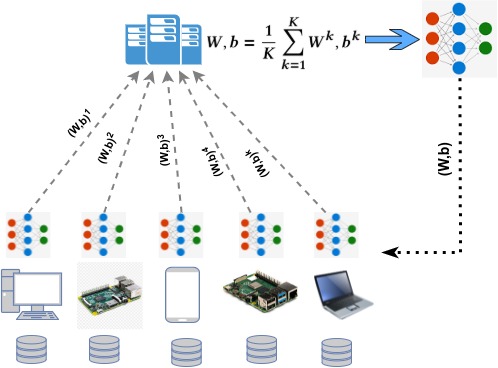}
\caption{ Decentralized federated learning training paradigm  }
\label{fig:learning}
\end{figure}

For every participating device, it is a machine learning task where it trains local model parameters using its data $D_i$. It takes input data $(x_j, y_j)_{j=1}^n$  to compute local parameters i.e. weights and biases. The loss function of every devices $i$ for dataset $D_i$ is \( F_i(w) =  \frac{1}{|D_i|} \sum\limits_{j=1}^n f(h(w,x_j),y_j) \). Where $f(h(w,x_j),y_j)$ is the loss for $j^{th}$ sample from $D_i$. The participating devices optimize the loss using an optimizer to find optimal parameters. In the subsequent step, the locally trained parameters (weights and biases) are shared with the central server for global model creation. The central curator receives all $k$ locally trained models parameters and performs aggregation operations on them. The weights and biases ($W,b$) of respective layers of every model are aggregated. Thus, the aggregated model contains representation from all models, which is further fine-tuned in the next iteration. The global model training paradigm is done in collaboration with central serve as shown in Fig~\ref{fig:learning}. The aggregated/updated global parameters ($W,b$) are pushed back to local devices for the next round of training. It is an iterative process that optimizes global parameters using local model updates.  The global loss function for the system is  \( F(w) = \frac{1}{|D|} \sum\limits_{i=1}^k |D_i| \times F_i(w)\). The goal of federated learning is to learn a global model by combining all local models. This training cycle continues until convergence without accessing raw data as shown in Fig~\ref{fig:learning}.

\subsection{Architecture}

\begin{figure}[h]
\centering
\includegraphics[width=\textwidth]{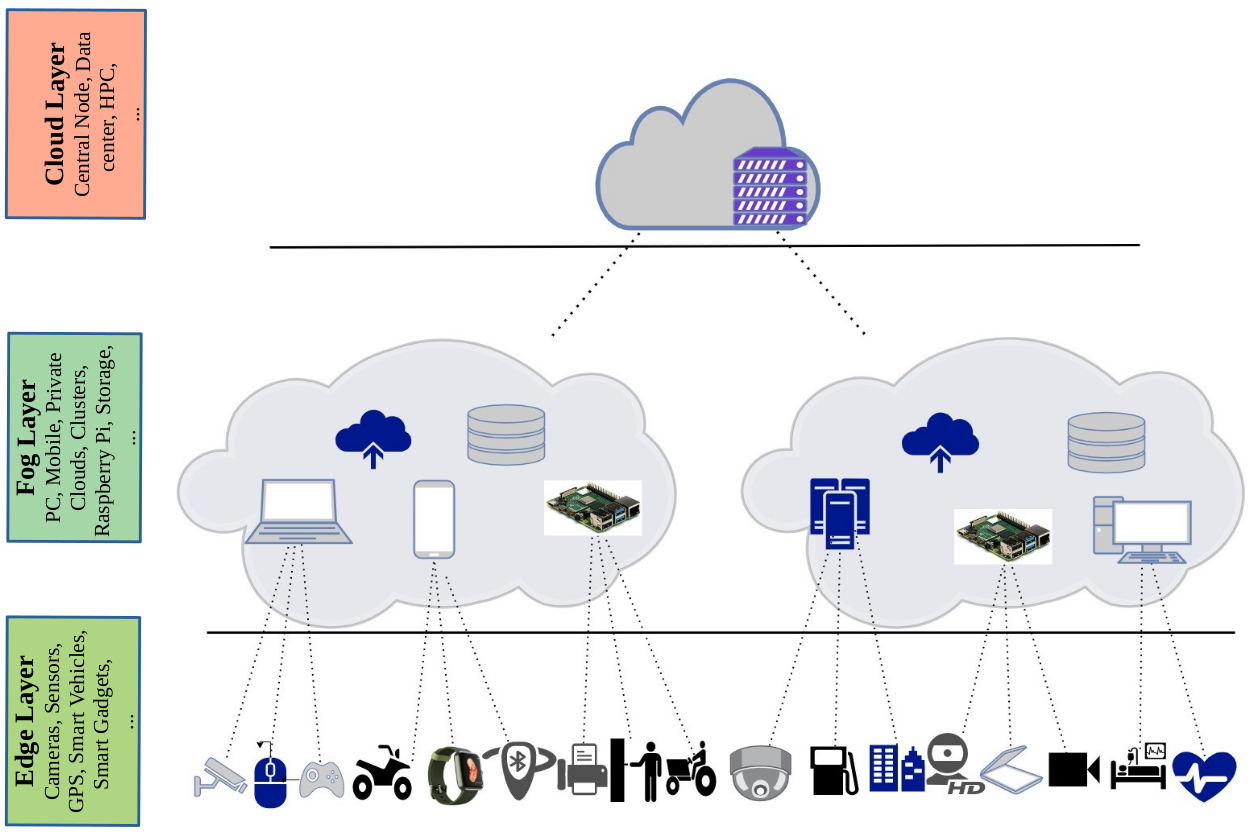}
\caption{ Distributed learning architecture  }
\label{fig:architecture}
\end{figure}

Sending raw data to the server consumes large bandwidth and creates network congestion. So it is not recommended for systems such as real time, delay-sensitive. Further, it also has privacy concerns. Another way is to compute the model on edge devices in collaboration with the cloud that suffers from high latency. However, the IoT devices have low computational resources that can not train machine learning model. To address this, we propose a fog enabled cloud-IoT online training architecture to simulate a machine learning model from continuously generated data by various IoT devices efficiently. The IoT nodes generate continuous data and share it with the fog node. A fog node has sufficient computational resources to train a machine learning model. It also has storage capability to store historical data generated from IoT devices. The cloud server facilitates global model creation by aggregating all participating devices leanings. The proposed architecture is shown in Fig \ref{fig:architecture}.

The Edge layer contains a large number of resources constrained IoT devices such as cameras, watches, GPS, bulbs, sensors, radars, etc. These devices continuously generate raw data by sensing the surroundings but are limited in storage and computations. The edge layer directly connects with the fog layer and share their data to the associated fog node. With enough computational power, a fog node trains a machine learning model in collaboration with the central server. Although, a fog node stores historic data of associated devices but the training is done on recently captured periodic data frames.  This simulates online training of continuously changing datasets. The Fog-Cloud layers participate in federated learning for machine learning. Once local models are trained on the fog layer, it is shared with the cloud layer. The cloud layer aggregates the local models and creates a global model. The federated learning with fog-cloud architecture is continued till the convergence of the global model. The proposed architecture is used for machine learning model training on rapidly growing data on resource constrained devices. The fog layer is responsible for data collection and federated learning training with a cloud node. While federated training fog layer only shares learning parameters to the cloud for aggregations. The raw data is stored at the fog layer, which is not shared with the central server. The proposed architecture simulates a distributed machine learning  using computations of resource constrained devices. 

\subsection{Online training and Data privacy}

With the variety of digital devices, data proliferation is another challenge in machine learning training. As discussed earlier,  sending this data to the server would not be an efficient way to mine it. Additionally, collecting, storing all the data, and then applying machine learning training needs huge computational resources. The data’s velocity increases as the number of devices employed increases. Consequently, it increases volume of the data with a variety of data that lead to big data problems. To address this to some extent, a continuous learning approach is applied while training a model. Rather than applying training on the complete dataset, online training is done on a subset of data. The subset can be periodic data generated from a device over a fixed time interval. Once the device learns the representation from current data, it shares representations with the central curator for global modelling. The curator aggregates all the representations and creates a central model for all devices. In the next round, every device generates a new set of datasets. Subsequently, in this round, the global model is further fine-tuned with the next dataset. This is important because IoT devices are low resources devices that are incapable of machine learning training on a very large dataset. The devices continue to generate the raw data and train the model. This can simulate a global model by exploiting low constrained devices computations.

However, the IoT devices such as sensors, CCTV cameras, radars do not have enough computation resources to run on-device machine learning. This work addresses this issue by deploying a fog node near IoT devices. All data generating devices are connected to the associated fog node to share raw data. With sufficient computational and storage capabilities, a fog node stores periodic data and performs online machine learning training. The rapid growth of data accumulates a large amount on a fog node. Due to the computational limitation of a fog node, online training is done on recently captured data leaving historic data aside. At the same time, the entire data is kept at the fog layer on backup storage, which can be reproduced in future if required. Then fog node participates in the federated learning process in collaboration with the central cloud server.    

IoT devices contain private data, location information, sensitive data, bank details, chat and personally identifiable data. Data privacy and security is another major challenge in cloud-IoT computation. Nowadays, there are increasing concerns for personal data sharing. The users are not comfortable in sharing personal data such as photos or chat to the cloud. The raw data contains crucial patterns that can be useful for various applications such as recommender systems, security analyses, smart homes, safety predictions, etc. Additionally, machine learning task has to follow strict data protection rules such as General Data Protection Regulation (GDPR). To address the privacy concerns, we have used a decentralized machine learning approach for model training.  In the proposed work, the data is stored at the fog layer and not shared with the cloud. Fog node participates in machine learning training that only shares model parameters, not raw data. This makes the system overall privacy preserving.

\section{Evaluation and results}
\label{sec:Evaluation}

This section describes evaluations and experimental results of the proposed framework for machine learning model. The model training is done for radar data in IIoT setup. We simulate the proposed architecture using Docker containers. Then, training of a global model for safe distance detection is done for human position in HR workspace. To show the efficiency of the proposed work, We have trained the ANN model in distributed environments and achieved expected results. 

\subsection{Docker based fog federation framework %for the evaluation
}

We have used docker containers to simulate the distributed machine learning. The docker engine facilitates multiple containers to run various programs independently. A container provides a run time environment for program execution. Additionally, docker creates a network of multiple containers that can communicate to others. With sufficient computational resources, we employed multiple docker containers as a fog node.  Every container runs a machine learning model independently with their local data. We have used gRPC library for requests and service calls between fog and central node. The federated learning is done between cloud and fog nodes using docker containers with gRPC calls. The IoT devices generate continuous data and share it with the fog node. Fog node stores historical data at backup devices and trains the model on recent data. To simulate the continuous data generation and online training, we used a fixed set of data samples for local training. Every container has its personal data, and machine learning is done on fixed periodic sequential training samples. 

We have simulated the proposed architecture to train ANN based machine learning model for human operator position detection in a human-robot workspace. The dataset is recorded from multiple Frequency-Modulated Continuous Wave (FMCW) radars. The fog nodes compile one minute of data from every device and complete one round of federating learning. We trained the model on 60 frames assuming every radar is generating 1 frame/sec. The next round of training is done on the next sequence of datasets. For this experiment, we have taken 5 fog nodes for decentralized machine learning training. The fog node trains fully connected ANN using tensorlow framework. The ANN model contains a single hidden layer with 64 neurons. The input and output layers have 512 and 8 neurons, respectively, based on data dimension and output labels.

We have trained a shallow neural network with one hidden layer. The model is a fully connected dense network with 64 units in the hidden layer. Input layer has 512 input neurons which is fully connected to the only hidden layer (dimension 512$\times$64 + 64 bias) followed by a ReLu layer. The output layer has 8 neurons which is densely connected to the hidden layer (dimension 64 $\times$ 8 + 8 bias). The final output label is predicted based on softmax activation function at the output layer. The local training on every device is done for 5 epochs to simulate low computational resources. The network is trained by backpropogation algorithm using categorical crossentropy loss function. Further, the 'adam' optimizer is used with learning 0.001 as an optimizer to optimize the training error. The loss value of the global model is calculated on test data. Also, we have traced the accuracy performance of the global model on both personal and unknown test datasets. The fog node participates in federated learning in collaboration with a central node with its local data, where device local data is recently generated one minute data. The simulation is done to show computational intelligence of proposed work on continuously changing data.

\subsection{FMCW radar dataset for federated learning}
The proposed architecture is validated on a real world IIoT use case. The data is generated from FMCW radars in a human-robot workspace. FMCW radars are effective IIoT device in industrial setup for environment sensing, distance measurement, etc. These radars are placed in a shared workspace of human robot to capture human position in the environment. Detection of human position in an industrial setup is crucial for worker's safety. These radars contain data to measure the distance and position of the human operator near it. The data distribution of devices is non independent and identically Distributed(non-IID), i.e. every device has its locally generated dataset. However, each participating devices have all classes samples. We used mentioned dataset to train a machine learning model to classify human safe distance. The dataset is published by Stefano Savazzi, which can be downloaded from IEEE Dataport~\cite{8yqc-1j15-19}. The radars output signals are preprocessed and converted into 512-point. The detailed methodology and collection of data are given in this paper~\cite{8950073}. 

\begin{table}[]
\centering
\caption{FMCW radars dataset}
\label{tab:dataset}
\begin{tabular}{|l|l|l|}
\hline
\textbf{Distance(m)}             & \textbf{Class} & \textbf{Critical/Safe} \\ \hline
\textless 0.5                    & 1              & Critical               \\ \hline
0.5 \textless{}= d \textless 1.0 & 2              & Critical               \\ \hline
1.0 \textless{}= d \textless 1.5 & 3              & Critical               \\ \hline
1.5 \textless{}= d \textless 2.0 & 4              & Safe                   \\ \hline
2.0 \textless{}= d \textless 2.5 & 5              & Safe                   \\ \hline
2.5 \textless{}= d \textless 3.0 & 6              & Safe                   \\ \hline
3.0 \textless{}= d \textless 3.5 & 7              & Safe                   \\ \hline
\textgreater{}=3.5               & 0              & Safe                   \\ \hline
\end{tabular}
\end{table}

The input sample contains 512 points with 8 labels. The labels are characterized by different distances of human operator and radar. The dataset contains a total of 32,000 samples of FFT range measurements 521 points. The dataset is already divided into training and testing samples of 16,000 $\times$ 512 shape. Additionally, the data sample is also randomly distributed over various devices for federated learning simulation in the database. We have implemented the given data distribution over 5 devices for federated learning to learn ANN model for safe/unsafe position detection.  The training is done for C=8 classification of the potential situation in human robot workspace. The label is an integer from 0 to 7, where Class 0 represents human distance  \textgreater 3.5m which is also marked as safe. Class 1 is represented as critical since the distance is \textless 0.5m. Other labels are marked based on different distance measures between humans and radars. Table~\ref{tab:dataset} contains labels for various classes.

\subsection{Results and Analysis}

\begin{figure}[h]
  \begin{subfigure}[b]{0.5\textwidth}
    \includegraphics[width=\linewidth]{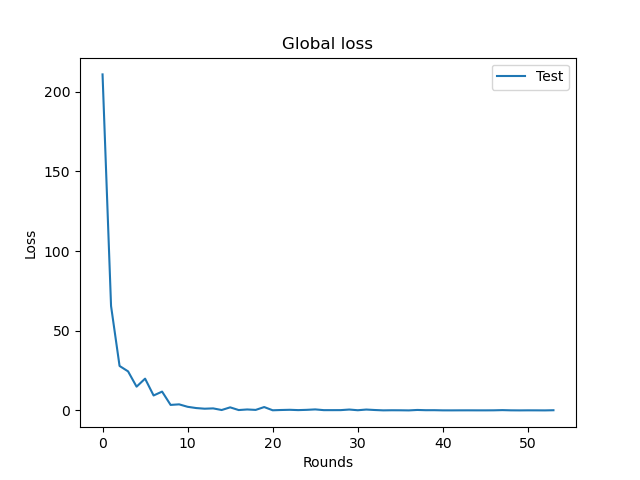}
    \caption{ }
    \label{fig:loss}
  \end{subfigure}
  \hfill
  \begin{subfigure}[b]{0.5\textwidth}
    \includegraphics[width=\textwidth]{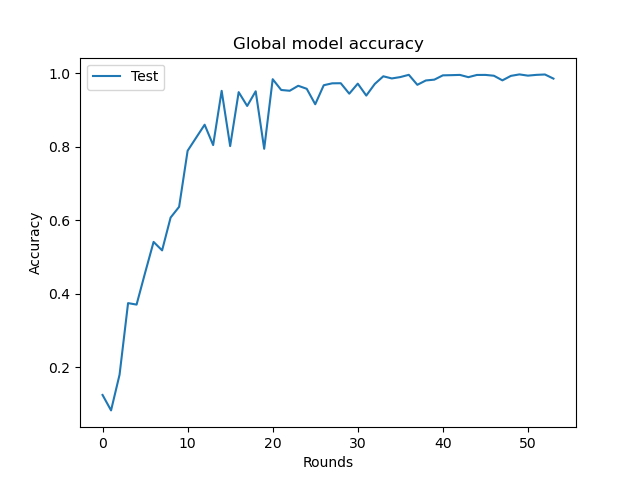}
    \caption{ }
    \label{fig:accuracy}
  \end{subfigure}
  \caption{Global model performance on test data: Loss value (a) and Accuracy (b) at every training round}
  \label{fig:performance}
\end{figure}

We trained an ANN model for human position classification in the shared HR workspace. The online training is done with 60 frames at a time with 5 fog nodes. From 16,000 training samples, every fog node receives 3200 independent samples. The local model training is done with the current 60 samples for 5 epochs only. Then central server performs aggregation of all learnt model parameters. This completes one round of federated learning. In the next round, we use next 60 samples for training by skipping previous data points. We executed such 53 rounds that exhaust entire local dataset training. At every round, we assess the model performance in terms of loss and accuracy. The global model is evaluated on the test dataset. The training loss and accuracy of the model on test data are shown in Fig~\ref{fig:performance}.

\begin{figure}[h]
\centering
\includegraphics[width=\textwidth]{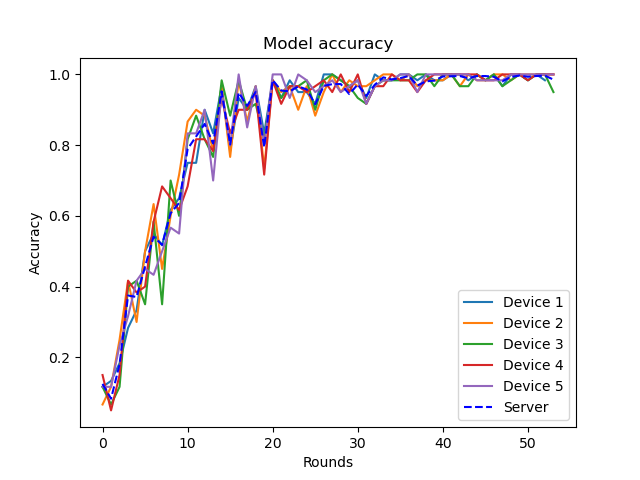}
\caption{ Model performance on all devices }
\label{fig:local_accuracy}
\end{figure}

As training increases over the number of rounds, model improves its accuracy significantly.  The model optimizes loss value and stabilizes the training after 30 rounds. The proposed online training performed exceptionally well as the test accuracy reached 99\%. The local model is trained on multiple fog nodes parallelly. The global model is combined learning of all local models. Fig~\ref{fig:local_accuracy} shows accuracy of the global model on various local data resides in fog devices. The accuracy of test data is averaged by cancelling the drift on various local training data. 

Training a machine learning model with a neural network is prone to overfitting, specifically in online training mode. Comparatively, a large model with huge parameters can memorize the data labels that tend to perform poorly on test data. In our experiment, the amount of data passed at a particular instance is relatively small (60 frames/iteration). The ANN model quickly learnt the sample with very high accuracy (\textgreater90\%) in fewer epochs. However, it failed to perform similar results on global test data. This is because of possible model overfitting on relatively small data. The federated learning consolidates such a few models to combined learnt information. The combined output model is better generalized and reduces possible overfitting. With the varying number of participating devices, federated learning prevents overfitting intrinsically. However, other generalization techniques such as dropout, normalization, regularization, etc may be applied at the architectural level.  This type of learning paradigm can help in creating a better generalized model that can be scope for future work.

\section{Conclusions and future work}
\label{sec:conclusion}

This work focuses on machine learning model training on decentralized data. We have proposed fog enabled distributed training architecture to train ML model on rapidly changing data. The architecture suitably uses decentralized algorithms such as federated learning for model creation. The edge layer is responsible for data generation. The cloud layer coordinates with computational nodes on the fog layer for machine learning. Whereas, the fog layer participates in distributed machine learning training with the central server. We have tested the proposed architecture on real world IIoT use case. The simulation result of position detection model trained on changing dataset is significant. We will further investigate the distributed architecture for communication and energy efficient training. Moreover, we only share trainable parameters to the server, not raw data. However, trainable parameters are vulnerable to attack. The robust privacy sensitive training could be another scope of work.

\section*{Acknowledgment}
We acknowledge financial support to UoH-IoE by MHRD, India (F11/9/2019-U3(A)).

\bibliography{elsarticle-template}

\end{document}